\documentclass[%
reprint,
superscriptaddress,
showpacs,preprintnumbers,
nofootinbib,
amsmath,amssymb,
aps,
prd,
floatfix,
]{revtex4-1}

\usepackage{graphicx}
\usepackage{dcolumn}
\usepackage{bm}
\usepackage{hyperref}
\usepackage[mathlines]{lineno}
\usepackage{float}
\usepackage{color}

\begin{document}

\preprint{IPPP/16/40, DCTP/16/80, IFT-UAM/CSIC-16-044}

\title{How to calculate dark matter direct detection exclusion limits that are consistent with gamma rays from annihilation in the Milky Way halo}

\author{David G. Cerde\~{n}o}
\email{davidg.cerdeno@gmail.com}
\affiliation{Institute for Particle Physics Phenomenology, Department of Physics
Durham University, Durham DH1 3LE, United Kingdom}
\affiliation{Instituto de F\'{\i}sica Te\'{o}rica UAM/CSIC, Universidad Aut\'{o}noma de Madrid, 28049, Madrid, Spain}

\author{Mattia Fornasa}
\email{fornasam@gmail.com}
\affiliation{GRAPPA Institute, University of Amsterdam, Science Park 904, 1098 XH Amsterdam, The Netherlands}

\author{Anne M. Green}
\email{anne.green@nottingham.ac.uk}
\affiliation{School of Physics and Astronomy, University of Nottingham, University Park, Nottingham, NG7 2RD, United Kingdom}

\author{Miguel Peir\'{o}}
\email{mpeirogarcia@gmail.com}
\affiliation{Instituto de F\'{\i}sica Te\'{o}rica UAM/CSIC, Universidad Aut\'{o}noma de Madrid, 28049, Madrid, Spain}
\affiliation{Departamento de F\'{\i}sica Te\'{o}rica, Universidad Aut\'{o}noma de Madrid, 28049 Madrid, Spain}

\date{\today}

\begin{abstract}
When comparing constraints on the Weakly Interacting Massive Particle (WIMP) properties from direct and indirect detection experiments it is crucial that the assumptions made about the dark matter (DM) distribution are realistic and consistent. For instance, if the Fermi-LAT Galactic centre GeV gamma-ray excess was due to WIMP annihilation, its morphology would be incompatible with the Standard Halo Model that is usually used to interpret data from direct detection experiments. In this article, we calculate exclusion limits from direct detection experiments using self-consistent velocity distributions, derived from mass models of the Milky Way where the DM halo has a generalized NFW profile. We use two different methods to make the mass model compatible with a DM interpretation of the Galactic centre gamma-ray excess. Firstly, we fix the inner slope of the DM density profile to the value that best fits the morphology of the excess. Secondly, we allow the inner slope to vary and include the morphology of the excess in the data sets used to constrain the gravitational potential of the Milky Way. The resulting direct detection limits differ significantly from those derived using the Standard Halo Model, in particular for light WIMPs, due to the differences in both the local DM density and velocity distribution.
\end{abstract}

\maketitle

\section{Introduction}
Weakly Interacting Massive Particles (WIMPs) are a well-motivated dark matter (DM) candidate. They can be detected directly, via their interactions with nuclei, or indirectly, via their annihilation products, such as high-energy gamma rays, neutrinos or anti-matter (for reviews see, e.g., Refs.~\cite{Jungman:1995df,Cerdeno:2010jj,Bertone:2004pz}).  The signals expected in these channels depend on different aspects of how DM is distributed in the Universe: the event rate in direct detection experiments depends on the local DM density and velocity distribution, while indirect searches are sensitive to non-local quantities. In particular, the gamma-ray flux expected from DM annihilations in a given direction in the sky is proportional to the line-of-sight integral of the square of the DM density towards that direction. For this reason, when comparing results from those two channels, it is  essential that the assumptions made about the DM distribution are consistent. It is also important  that the modelling of the DM distribution is as realistic as possible.

When direct detection data are used to constrain the WIMP mass and scattering cross-section, it is customary to assume the so-called Standard Halo Model (SHM). The SHM postulates a spherical DM halo for the Milky Way (MW) with a density profile that scales as $\rho \propto r^{-2}$, and an isotropic Maxwell-Boltzmann speed distribution \cite{Lewin:1995rx}. However, this is in conflict with numerical simulations, which produce DM halos that are not exactly spherical and have speed distributions that can deviate systematically from the Maxwell-Boltzmann distribution \cite{Hansen:2005yj,Vogelsberger:2008qb,Kuhlen:2009vh,Ludlow:2011cs,Mao:2012hf,Kuhlen:2013tra,Butsky2015,Bozorgnia:2016ogo,Kelso:2016qqj,Sloane:2016kyi}. Also, simulated halos have density profiles with logarithmic slopes which vary with radius~\cite{Navarro:1995iw,Navarro:1996gj, Graham:2005xx,Springel:2008cc,Navarro:2008kc}, in contrast to the constant logarithmic slope of the SHM.

A more complete description of the MW relies on modelling its matter components, i.e., the disk, bulge and DM halo. The free parameters of such a mass model can be constrained by means of astrophysical observables, such as the Oort constants, the local surface density or the microlensing optical depth  \cite{Bahcall:1980fb,Dehnen:1996fa,Ollig:2000,Klypin:2001xu,Catena:2009mf,Catena:2011kv,McMillan:2011wd,Fornasa:2013iaa}. For spherically symmetric halos, this approach allows the reconstruction of the phase-space distribution of DM particles in the MW, $F(\mathbf{x},\mathbf{v})$. From the latter it is possible to extract a self-consistent DM velocity distribution, $f(\mathbf{v})$, that is in agreement with the inferred gravitational potential of our Galaxy \cite{Fornasa:2013iaa}.

In this article, we extend previous studies by including indirect detection data from gamma-ray searches as a further constraint in the mass modelling of the MW. 
As a case study, we consider the excess of GeV gamma rays observed close to the Galactic centre~\cite{Vitale:2009hr,Goodenough:2009gk,Hooper:2010mq,Hooper:2011ti,Abazajian:2012pn,Gordon:2013vta,Macias:2013vya,Abazajian:2014fta,Daylan:2014rsa,Murgia:2014,Calore:2014xka,Calore:2014nla,Calore:2015nua,Porter:2015uaa,Linden:2016rcf} by the Fermi Large Area Telescope (LAT) as a potential DM signal. The energy spectrum, morphology and overall normalisation of this excess are consistent with the expectations for WIMP annihilations. In particular, the excess is roughly spherically symmetric and the morphology is consistent with a DM density profile proportional to $r^{-\gamma}$, with $\gamma \approx 1.1-1.3$~\cite{Daylan:2014rsa,Calore:2014xka}. However, the origin of the excess is not yet established and alternative astrophysical explanations including cosmic-ray outbursts~\cite{Carlson:2014cwa,Petrovic:2014uda,Cholis:2015dea}, a population of unresolved millisecond-pulsar-like sources~\cite{Abazajian:2010zy,Gordon:2013vta,Abazajian:2014fta,Yuan:2014rca,Bartels:2015aea,Lee:2015fea,Calore:2015bsx}, or additional cosmic-ray sources~\cite{Gaggero:2015nsa,Carlson:2016iis} have been considered. If interpreted in terms of DM annihilations, the morphology of this excess can determine, up to a certain degree of accuracy, the inner slope of the DM profile, thereby adding a valuable piece of information to the modelling of the gravitational potential of the MW.

Using the MW gravitational potential, we derive the self-consistent local speed distribution and local DM density. This allows the computation of the DM event rate in direct detection experiments, and can be used to extract consistent exclusion limits on the DM-nucleon scattering cross section. In this article, we apply this prescription to calculate consistent exclusion limits from the LUX data \cite{Akerib:2013tjd}.
This procedure permits an unbiased comparison of experimental results, which is essential when applying experimental results to specific particle models for DM.

In Sec.~\ref{sec:mass_model} we describe our mass model of the MW, before outlining how we self-consistently calculate the DM local velocity distribution in Sec.~\ref{sec:fv}. We then derive the resulting exclusion limits for the LUX experiment in Sec.~\ref{sec:dd} and conclude in Sec.~\ref{sec:discussion}.

\section{Mass models of the Milky Way}
\label{sec:mass_model}
A mass model of the MW is a description of the different components of the gravitational potential of our Galaxy (see Refs.~\cite{Bahcall:1980fb,Dehnen:1996fa,Ollig:2000}). The parameters specifying the different components can be reconstructed by requiring a good fit to various data sets including stellar kinematics (such as rotation curves in the inner Galaxy or velocity dispersion of halo stars) and microlensing. See Refs.~\citep{Klypin:2001xu,Widrow:2008yg,Catena:2009mf,Catena:2011kv,Iocco:2011jz,McMillan:2011wd,Nesti:2013uwa,Burch:2013pda,Bozorgnia:2013pua}, among others.

In this work, we use the mass model defined in Ref.~\cite{Fornasa:2013iaa}. Four matter components are considered: 
\begin{itemize}

\item a stellar disk with density $\rho_{\rm d}(R,z) = (\Sigma_{\rm d} / 2z_{\rm d}) \exp(-R/R_{\rm d}) \, \mbox{sech}^2 (z/z_{\rm d})$, where $R$ is the distance from the Galactic centre, projected on the Galactic plane, and $z$ is the vertical distance from the Galactic plane;

\item a combined stellar bulge and bar, i.e., $\rho_{\rm bb}(x,y,z) = \rho_{\rm bb}(0) [s_{\rm a}^{-1,85} \exp(-s_{\rm a}) + \exp(-0.5 s_{\rm b}^2)]$, where $(x,y,z)$ are Cartesian coordinates, $s_{\rm a}^2 = [q_{\rm b}^2 (x^2 + y^2) + z^2] / z_{\rm b}^2$ and $s_{\rm b}^4 = (x^2 + y^2)^2 / x_{\rm b}^4 + (z/z_{\rm b})^4$;

\item interstellar gas, modelled as in Ref.~\cite{Moskalenko:2001ya};

\item a DM halo with a generalised Navarro-Frenk-White (NFW) profile:
\begin{equation}
\rho_{\chi}(r) = \rho_{\rm s} (r/r_{\rm s})^{-\gamma} (1+r/r_{\rm s})^{\gamma-3}.
\label{eqn:generalised_NFW}
\end{equation}  
The case $\gamma=1$ corresponds to the original NFW profile~\cite{Navarro:1996gj}, which provides a good fit to halos formed in simulations containing only DM particles. Adiabatic contraction, as baryons infall, could lead to a steeper inner profile (e.g. Ref.~\cite{Gustafsson:2006gr,SommerLarsen:2009me}). While hydrodynamical simulations have recently become precise enough to reproduce the observed properties of galaxies \cite{Kuhlen:2013tra}, the complexity of phenomena like stellar winds and supernovae feedback are still hard to model and their interplay with DM is not yet fully understood \cite{Tissera:2009cm,Pontzen:2011ty,Maccio':2011eh}. The morphology of the excess from the Galactic centre is consistent with a DM halo with $\gamma \approx 1.26$~\cite{Daylan:2014rsa,Calore:2014xka}.
\end{itemize}

Some parameters in the expressions above ($z_{\rm d}$, $q_{\rm b}$, $z_{\rm b}$ and $z_{\rm b}$) are fixed to values inferred from astronomical observations. All the other parameters ($\Sigma_{\rm d}$, $R_{\rm d}$, $\rho_{\rm bb}(0)$, $\rho_{\rm s}$ and $r_{\rm s}$) are allowed to vary.

In this work, we consider three mass models of the MW, which differ in the assumptions made regarding the inner logarithmic slope, $\gamma$, of the density profile of the DM halo:
\begin{itemize}
\item a standard NFW halo (dubbed, from now on, simply ``NFW'') with $\gamma=1$. This commonly used profile is inconsistent with the DM interpretation of the Galactic Centre excess. We use it to illustrate the effects of different assumptions for the distribution of DM in the MW on the exclusion limits derived from direct detection data; \\
\item a NFW halo with $\gamma=1.26$ (c.f. Ref.~\cite{Calore:2014nla}) which we refer to as ``generalised NFW ($\gamma=1.26$)''. Here we have enforced compatibility with a DM interpretation of the Galactic Centre excess by simply fixing the inner slope to the value which best fits the morphology of the excess; \\
\item a NFW halo with a free $\gamma$, referred to as ``generalised NFW (free $\gamma$)''. In this case we include the morphology of the Galactic Centre excess in the data used to constrain the mass model (as discussed in detail below). This is a more sophisticated way of ensuring compatibility. As well as including the uncertainty in the inner slope, it allows for possible degeneracies in the parameters of the mass model.
\end{itemize}
The last two cases were not previously considered in 
Ref.~\cite{Fornasa:2013iaa}.

\begin{figure*}
\begin{center}
\includegraphics[width=0.4\textwidth]{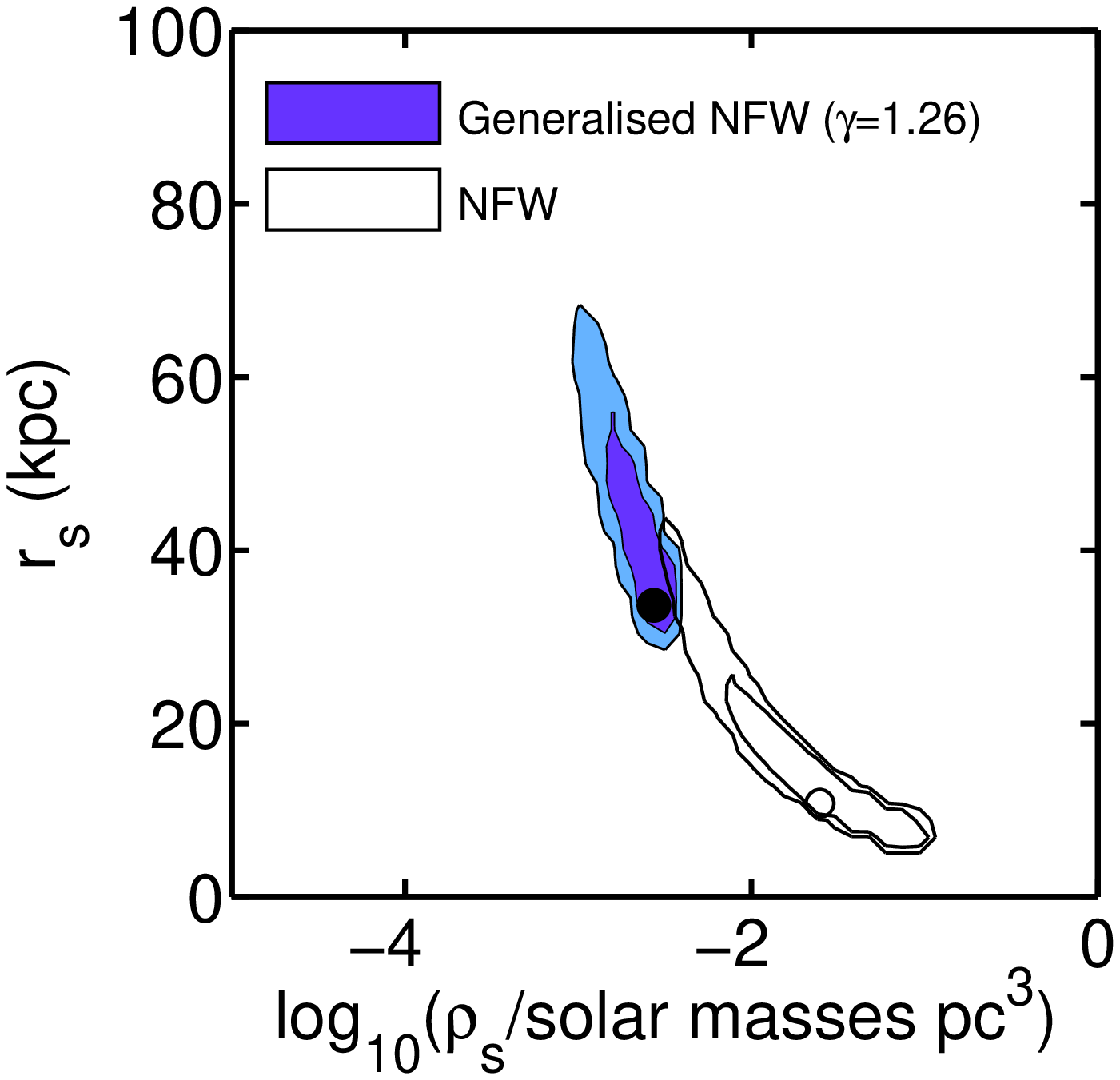}
\includegraphics[width=0.4\textwidth]{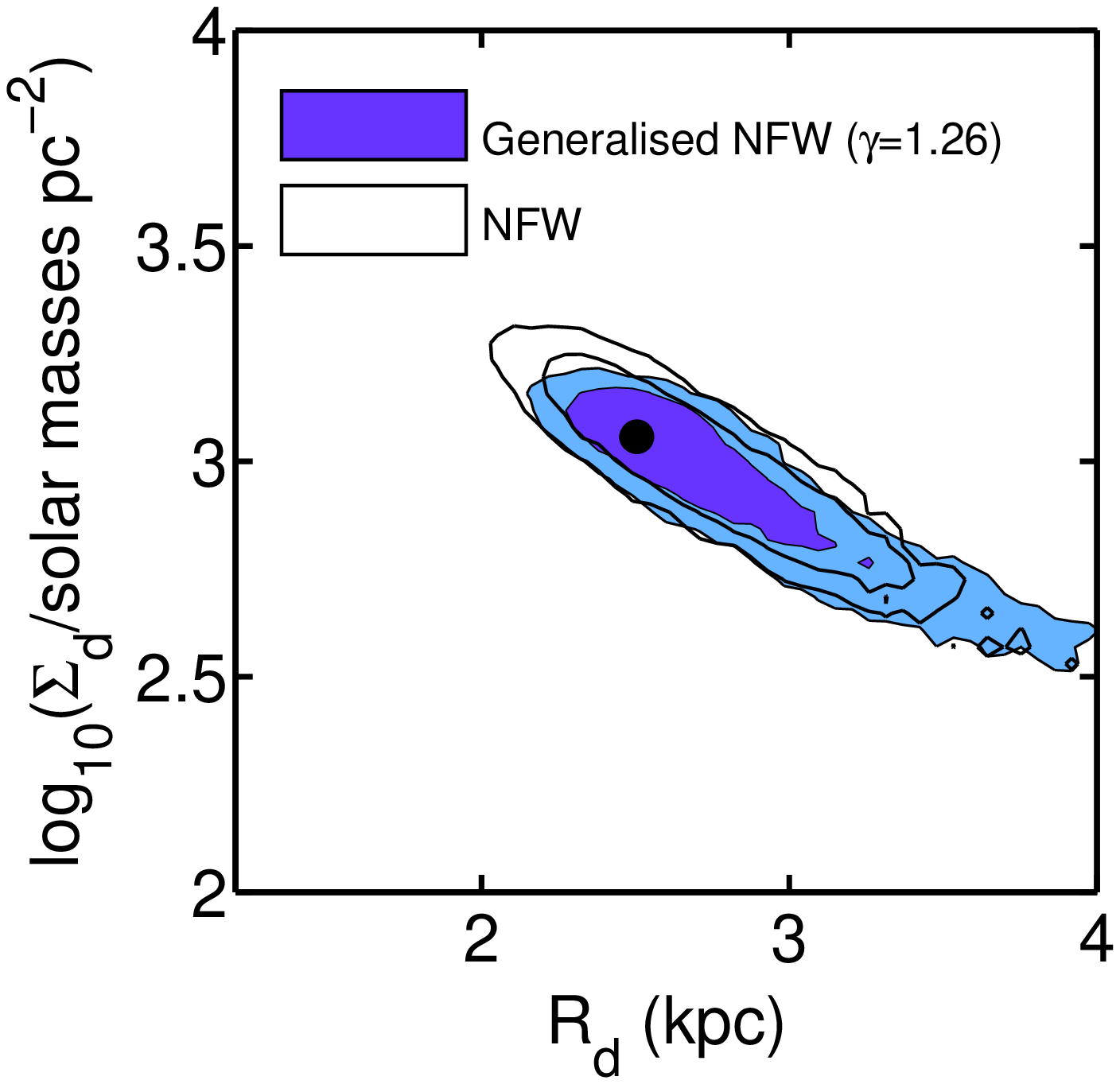}
\vspace*{-0.7cm}
\end{center}
\caption{\label{fig:NFWcf_mass_model} 2D probability distributions for the scale density, $\rho_{\rm s}$, and scale radius, $r_{\rm s}$, defining the MW DM halo (left) and for the surface density, $\Sigma_{\rm d}$, and the radius, $R_{\rm d}$, defining the disk (right). The coloured regions correspond to the ``generalised NFW ($\gamma=1.26$)'' case while the empty ones are for the ``NFW'' case. The inner and outer contours denote the 68\% and 95\% confidence regions for the profile likelihood. The full and empty dots indicate the best-fit points for the ``generalised NFW ($\gamma=1.26$)'' and ``NFW'' cases, respectively.}
\end{figure*}

We have performed Bayesian scans over the free parameters (10 for the ``NFW'' and ``generalised NFW ($\gamma$=1.26)'' cases and 11 for  ``generalised NFW (free $\gamma$)'') in order to determine the configuration that best fits a set of experimental data. These data have been chosen to maximise the precision in the reconstruction of the parameters of the mass model and, thus, of the MW gravitational potential. They are:
\begin{itemize}
\item the local circular velocity, $\Theta_0$, inferred from the motion of SgrA$^\star$ \cite{Reid:2004rd}, with the rotational component of the Sun's velocity with respect to the Local Standard of Rest taken from the analysis of 3000 stars in
the APOGEE survey \cite{Bovy:2012ba};
\item the sum of the Oort constants, measured from the motion of Cepheid stars from the Hipparcos satellite \cite{Feast:1997sb};
\item the local surface density within a distance of 1.1 kpc from the Galactic plane. The measurement is taken from Refs.~\cite{Kuijken:1989,Kuijken:1991};
\item the rotation curve between $0.35\,R_0$ and $0.9\,R_0$, where $R_0$ is the Solar radius, using terminal velocities obtained from the spectral line of atomic hydrogen HI in Ref.~\cite{Malhotra:1994qj};
\item the velocity dispersion of 2000 Blue Horizontal-Branch stars observed by the Sloan Digital Sky Survey. The data, analysed in Ref.~\cite{Xue:2008se}, cover distances between 5.0 and 60 kpc from the Galactic centre;
\item 10 measurements of microlensing optical depth from the MACHO \cite{Popowski:2004uv}, OGLE-II \cite{Sumi:2005da} and EROS collaborations \cite{Rahal:2009yt};
\item The morphology of the Galactic Centre excess. This data is included only for the ``generalised NFW (free $\gamma$)'' DM halo in order to constrain the mass model near the centre of the MW and also ensure compatibility with a DM interpretation of the Galactic centre excess. Fig.~1 of Ref.~\cite{Calore:2014nla} shows the morphology of the excess as a function of Galactic latitude. We consider three data points, namely the ratios between $i)$ the emission at $2.5^{\circ}$ and $7.5^{\circ}$, $ii)$ between $7.5^{\circ}$ and $12.5^{\circ}$ and, $iii)$ between $12.5^{\circ}$ and $17.5^{\circ}$ (in each case we use the fluxes at the centre of the pink bands in Fig.~1 of Ref.~\cite{Calore:2014nla}). We have used the ratios of the intensities rather than the actual fluxes, since the ratios fix the morphology of the signal (and, therefore, the parameters of the MW DM halo) irrespective of the DM annihilation cross-section and channel. For the ``NFW'' case the gamma-ray data are neglected. For the ``generalised NFW ($\gamma=1.26$)'' case, compatibility with a DM interpretation of the Galactic Center excess is instead enforced by setting $\gamma$ equal to 1.26 by hand.
\end{itemize}

\begin{figure*}
\begin{center}
\hspace*{-1cm}
\includegraphics[width=0.36\textwidth]{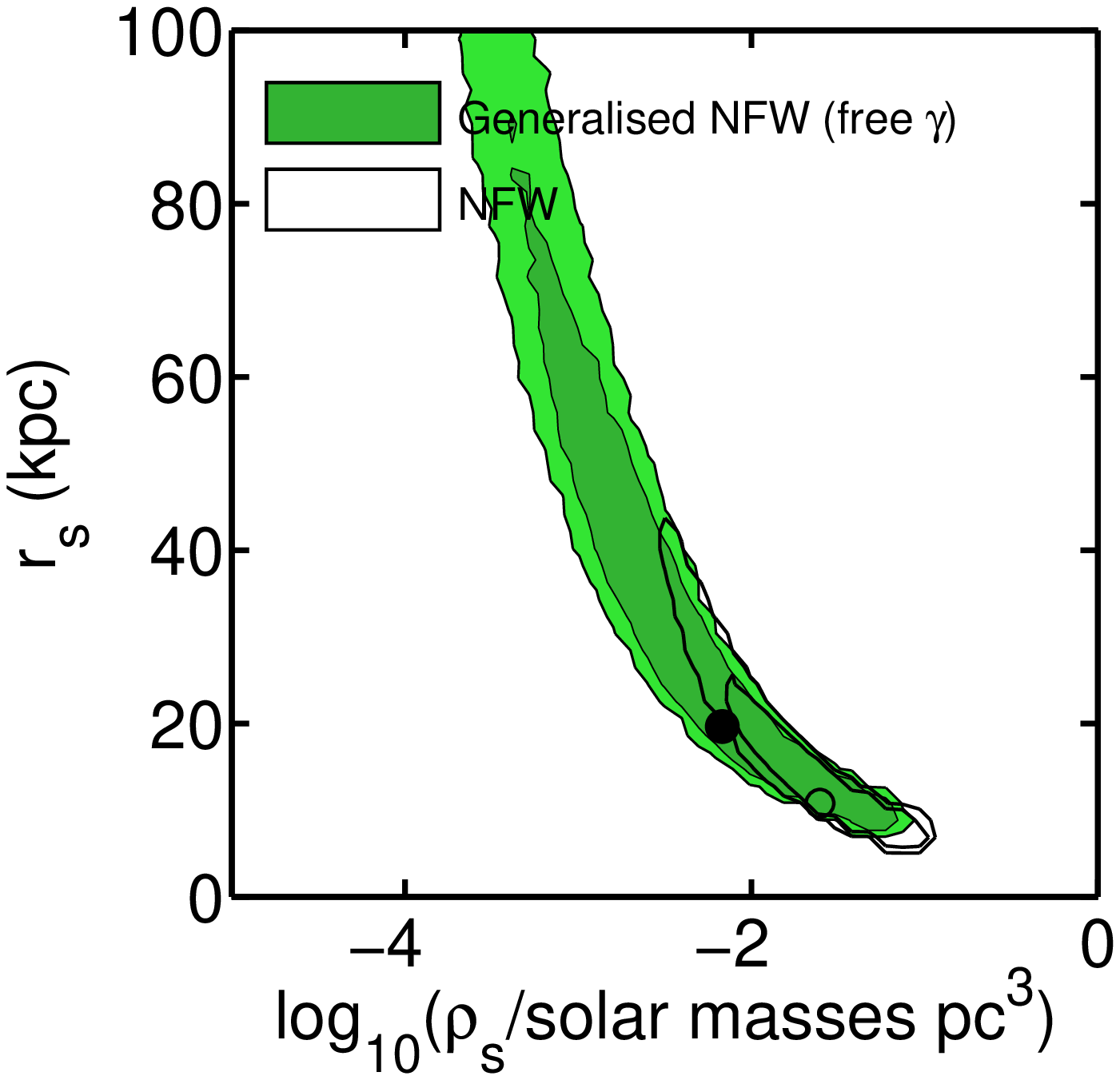}
\hspace*{-0.5cm}
\includegraphics[width=0.36\textwidth]{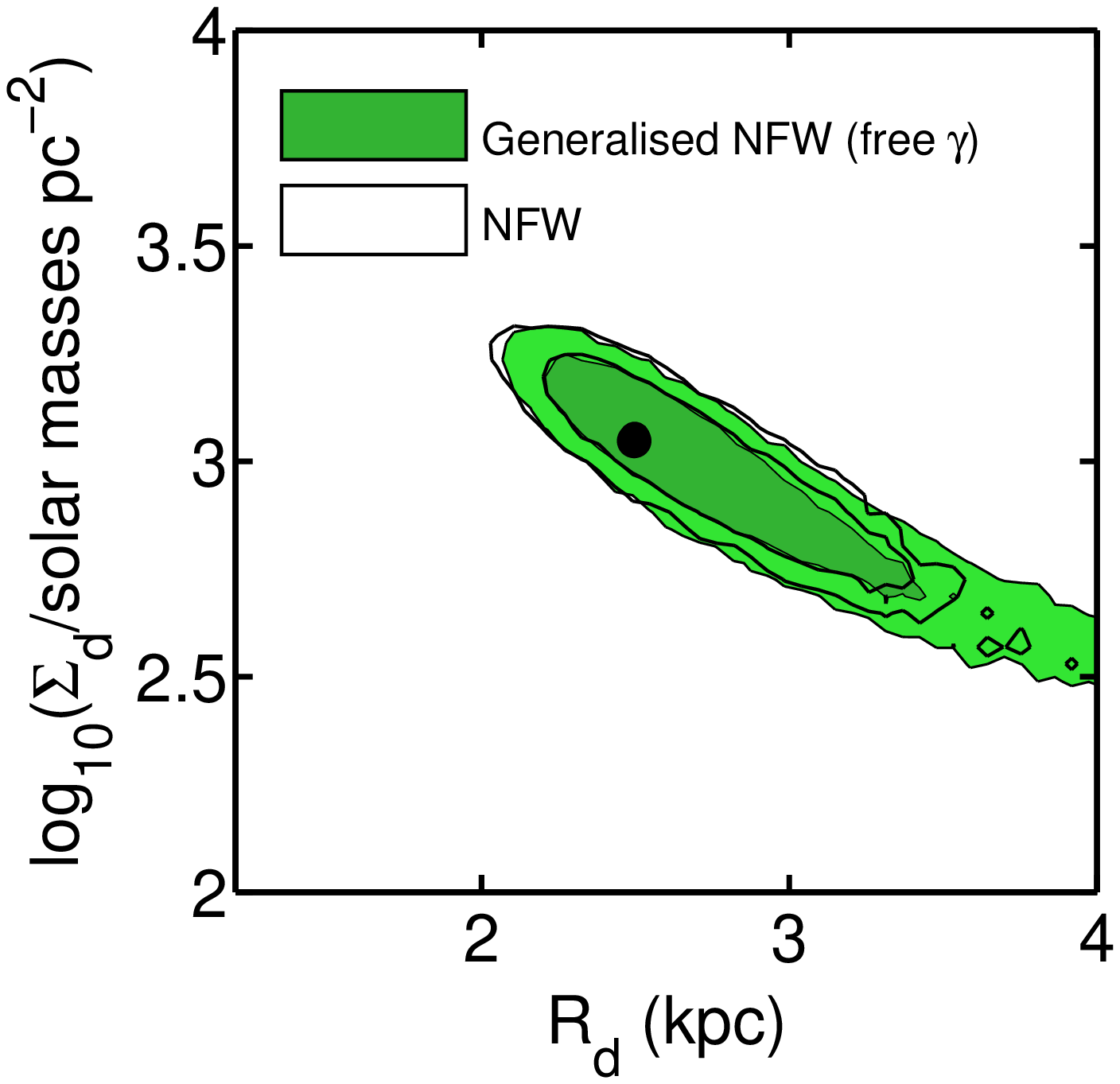}
\hspace*{-0.5cm}
\includegraphics[width=0.36\textwidth]{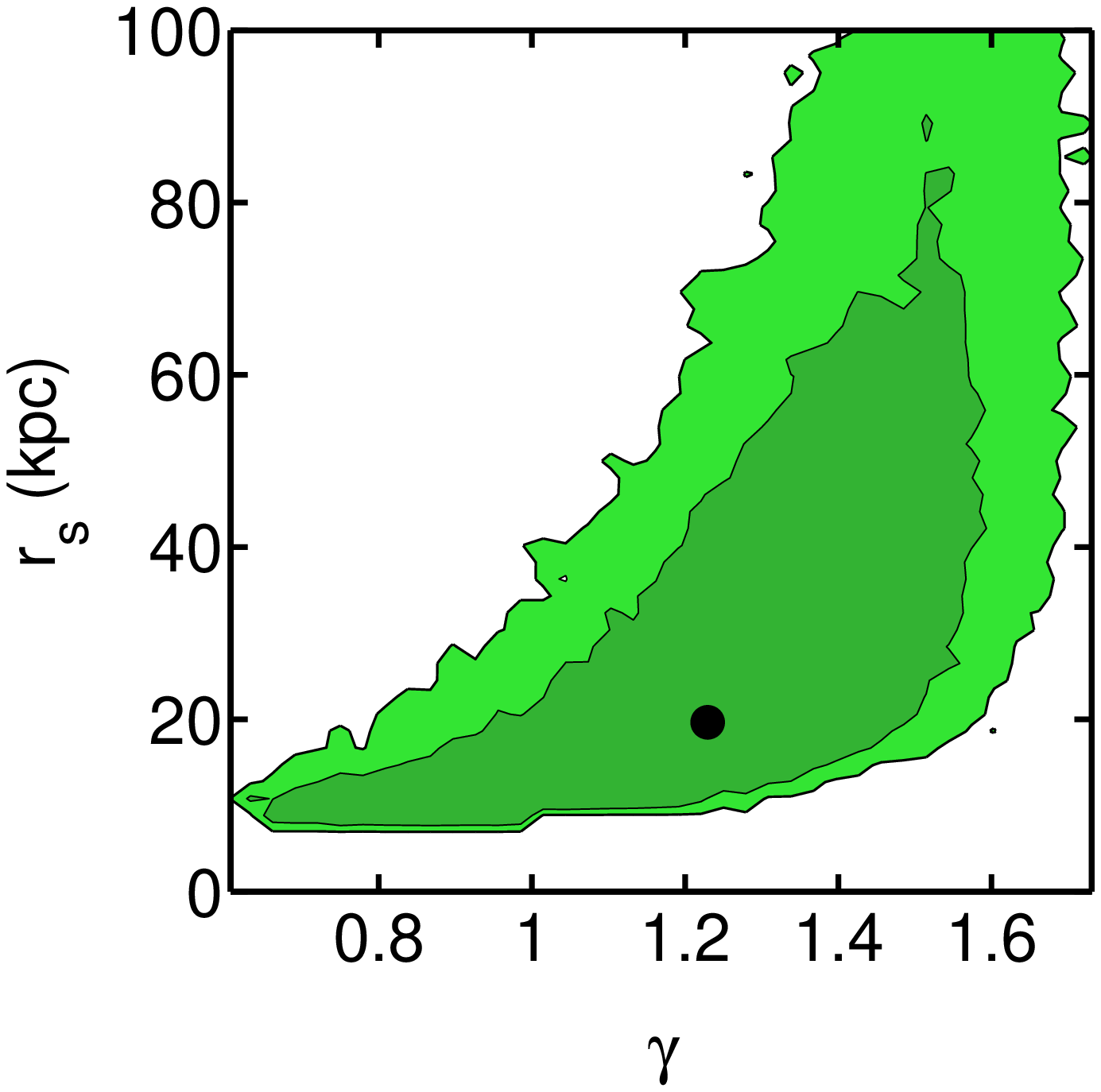}
\hspace*{-1cm}
\vspace*{-0.7cm}
\end{center}
\caption{\label{fig:NFWc_mass_model} Left and centre: the same as in Fig.~\ref{fig:NFWcf_mass_model}, but now the coloured contours correspond to the ``generalised NFW (free $\gamma$)" case. Right: 2D probability distributions of the DM halo scale radius, $r_{\rm s}$, and inner slope, $\gamma$, for the generalised ``NFW (free $\gamma$)" case.}
\end{figure*}

We have constructed a likelihood function, assuming that each data set is characterised by a Gaussian probability distribution, and used MultiNest v3.9 \cite{Feroz:2013hea} to carry out Bayesian scans to derive the probability distributions of the free parameters.

Fig.~\ref{fig:NFWcf_mass_model} shows the resulting 2D probability distributions for the profile likelihood of the parameters defining the MW DM halo ($\rho_{\rm s}$ and $r_{\rm s}$) and the disk ($\Sigma_{\rm d}$ and $R_{\rm d}$). The coloured regions are for ``generalised NFW (fixed $\gamma=1.26$)'', while the empty ones are for ``NFW''. Inner and outer contours indicate the 68\% and 95\% confidence regions, respectively. The full and empty dots mark the position of the best-fit points for the ``generalised NFW (fixed $\gamma=1.26$)'' and ``NFW'' cases, respectively. Lines of constant enclosed DM mass, $M_{\chi}(<r)$, in the plane $(\rho_{\rm s}, r_{\rm s})$ are hyperbolae. When $\gamma$ is increased from 1 (empty contours in Fig.~\ref{fig:NFWcf_mass_model}) to 1.26 (filled contours), the preferred region moves approximately along the same hyperbola, in order to reproduce the same value of the circular velocity and, thus, the same enclosed mass within $R_0$. However, for larger $\gamma$ the preferred region corresponds to a more extended halo, i.e., a larger $r_{\rm s}$. Indeed, the best-fit halo concentration~\footnote{The concentration $c_\Delta$ is defined as the ratio between the distance enclosing a density that is $\Delta$ times the critical density of the Universe and $r_{\rm s}$. In terms of the matter density parameter, $\Omega_{\rm m}(z)$, we use  $\Delta = 18 \pi^2 + 82 ( \Omega_{\rm m}(z) - 1) - 39 (\Omega_{\rm m}(z) - 1)^2$.}  $c_{\Delta}$ decreases from $23.2$ to $10.1$.

Fig.~\ref{fig:NFWc_mass_model} shows the same plots as 
Fig.~\ref{fig:NFWcf_mass_model}, but, in this case, the coloured contours correspond to the ``generalised NFW (free $\gamma$)'' case. An additional plot is included, showing the preferred region in the $(\gamma,r_{\rm s})$ plane. Leaving
$\gamma$ free to vary allows the preferred region to extend to large values of $r_{\rm s}$, corresponding to large values of $R_{\rm d}$. Note that the best-fit value for $\gamma$ is 1.23, but the 68\% credible region extends from $0.6$ to approximately $1.6$. The best-fit solution is closer to the "NFW" best-fit halo than in Fig.\,\ref{fig:NFWcf_mass_model}, and has concentration $c_\Delta=14.4$.

\section{Self-consistent velocity distributions}
\label{sec:fv}
The phase-space distribution, $F({\bf x},{\bf v})$, of a matter component is self-consistent if, together with the gravitational potential of the system, $\Phi({\bf x})$, it satisfies the Boltzmann equation (see, e.g., Refs.~\cite{Binney_Tremaine,Catena:2011kv,Fornasa:2013iaa}). For a spherically symmetric system with an isotropic velocity distribution, the phase-space distribution is a function of the binding energy, $E$, only. It is completely determined by the gravitational potential and can be calculated using the Eddington equation~\cite{Eddington:1915}.

For the more general case of a spherically symmetric system with an anisotropic velocity tensor, the phase-space distribution depends on the modulus of the angular momentum, $L$, as well. In this case a parametric form is often assumed for $F(E,L)$, and for some sets of parameters, self-consistent solutions can be found. We follow Ref.~\cite{Wojtak:2008mg} which assumes that the phase-space distribution is separable, $F(E,L)=F_{E}(E)F_{L}(L)$, a hypothesis that has been verified for simulated cluster-size DM halos~\cite{Wojtak:2008mg}. The authors of Ref.~\cite{Wojtak:2008mg} also assume that a reasonable parameterisation for $F_{L}(L)$ is given by the 
following expression:
\begin{equation}
F_L(L) = \left( 1 + \frac{L^2}{2L_0^2} \right)^{-\beta_\infty+\beta_0} L^{-2\beta_0} \,.
\label{eqn:F_L}
\end{equation}
Here $\beta(r)$ is the velocity anisotropy parameter  
\begin{equation}
\beta(r) = 1 - \frac{\sigma_{\rm t}^2}{2\sigma_{\rm r}^2} \,,
\end{equation}
where $\sigma_{\rm t}$ and $\sigma_{\rm r}$ are the tangential and radial velocity dispersions, and $\beta_{0}$ and $\beta_{\infty}$ are its values at $r=0$ and infinity, respectively. $L_{0}$ governs the transition of $\beta(r)$ between these values. 
The self-consistent solution obtained from the assumption in Eq.~(\ref{eqn:F_L}) matches the velocity anisotropy of simulated cluster-sized halos \cite{Wojtak:2008mg}, in which $\beta$ is zero close to the centre, growing to $\beta \sim 0.2$ at $r_{\rm s}$, and increasing further to $\beta \sim 0.4$ at $r \sim 10 \, r_s$. Similar behaviour has been found in simulations of MW-like halos~\cite{Ludlow:2010sy,Ludlow:2011cs}, although in this case, $\beta$ may decrease beyond $r \sim 5 \, r_s$. This behaviour can, however, also be reproduced by Eq.~(\ref{eqn:F_L}).

For a fixed gravitational potential and particular values of $L_{0}$, $\beta_{0}$ and $\beta_{\infty}$, $F_{E}(E)$ can be determined by inverting the following equation:
\begin{eqnarray}
\rho_\chi(r) & = & \int {\rm d}^3v \, F_E(E) F_L(L) \,,  \nonumber \\
& = & \int {\rm d}^3v \, F_E(E) 
\left( 1 + \frac{L^2}{2 L_0^2} \right)^{-\beta_\infty + \beta_0} L^{-2\beta_0}.
\label{eqn:Volterra}
\end{eqnarray}
This is a Volterra integral equation which has to be solved numerically
\cite{Wojtak:2008mg}.

Using the phase-space distribution function, we can now derive the
WIMP local speed distribution, $f_{1}(v)$. This is the relevant quantity for direct detection experiments, and is defined as
\begin{equation}
f_{1}(v) = \int v^2 f(\mathbf{v}) \, {\rm d}\Omega_{\mathbf{v}} = 
\frac{\int v^2 \, F(\mathbf{x}_\odot,\mathbf{v}) \, {\rm d}\Omega_{\mathbf{v}}}{\rho_\chi({\mathbf{x}_\odot})} \,.
\label{eq:fv_galframe}
\end{equation}
The speed distribution is equal to zero for speeds larger than the local escape velocity in the Galactic rest frame $v_{\rm esc}$. As discussed in Ref.~\cite{Fornasa:2013iaa}, the astronomical observations we use to constrain the gravitational potential do not provide any information on $L_{0}$, $\beta_{0}$ and $\beta_{\infty}$ and, hence, these parameters must be marginalized over. This can lead to large uncertainties in $F(E,L)$. However, the shape of the speed distribution is directly constrained by some of the observational data of Sec.~\ref{sec:mass_model}. For instance the distribution does not extend beyond the escape velocity (which is a measure of the local gravitational potential), while the position of the peak is related to the local circular velocity. Therefore, $f_{1}(v)$ can be reconstructed with reasonable uncertainties, even after the marginalisation of $L_0$, $\beta_0$ and $\beta_\infty$. The effect of $\beta_0$ is localised mainly at small velocities. The reason for this is that particles with small velocities have large binding energies (since $E= \Phi - v^2/2$) and, thus, correspond to orbits that are localised close to the centre of the halo. Conversely, particles with large velocities have small binding energies and can reach large radii, and hence become sensitive to $\beta_{\infty}$. When $L_{0}$ is small, increasing its value has the same effect as decreasing $\beta_0$. On the other hand, for $L_0 \sim R_0 \Theta_0$, the effect of increasing $L_0$ is the same as increasing $\beta_\infty$. Finally, for large $L_0$, the velocity distribution becomes independent of $L_0$ since further increasing the transition scale only affects large radii. See Ref.~\cite{Fornasa:2013iaa} for a detailed discussion of how $F_{E}(E)$ and $f_{1}(v)$ depend on both the gravitational potential and the velocity anisotropy of the system.

\begin{figure*}
\begin{center}
\includegraphics[width=0.4\textwidth]{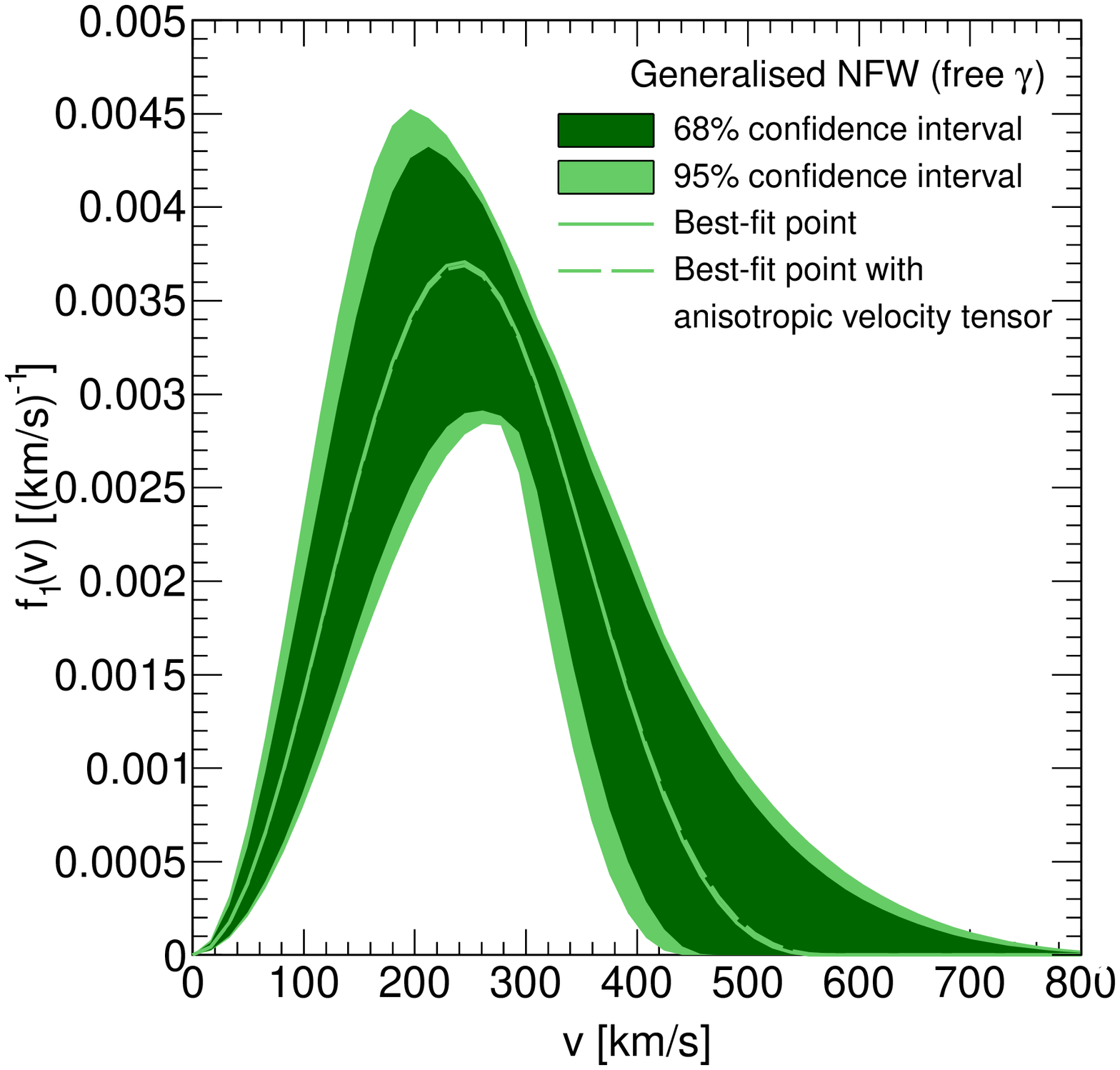} \\
\includegraphics[width=0.4\textwidth]{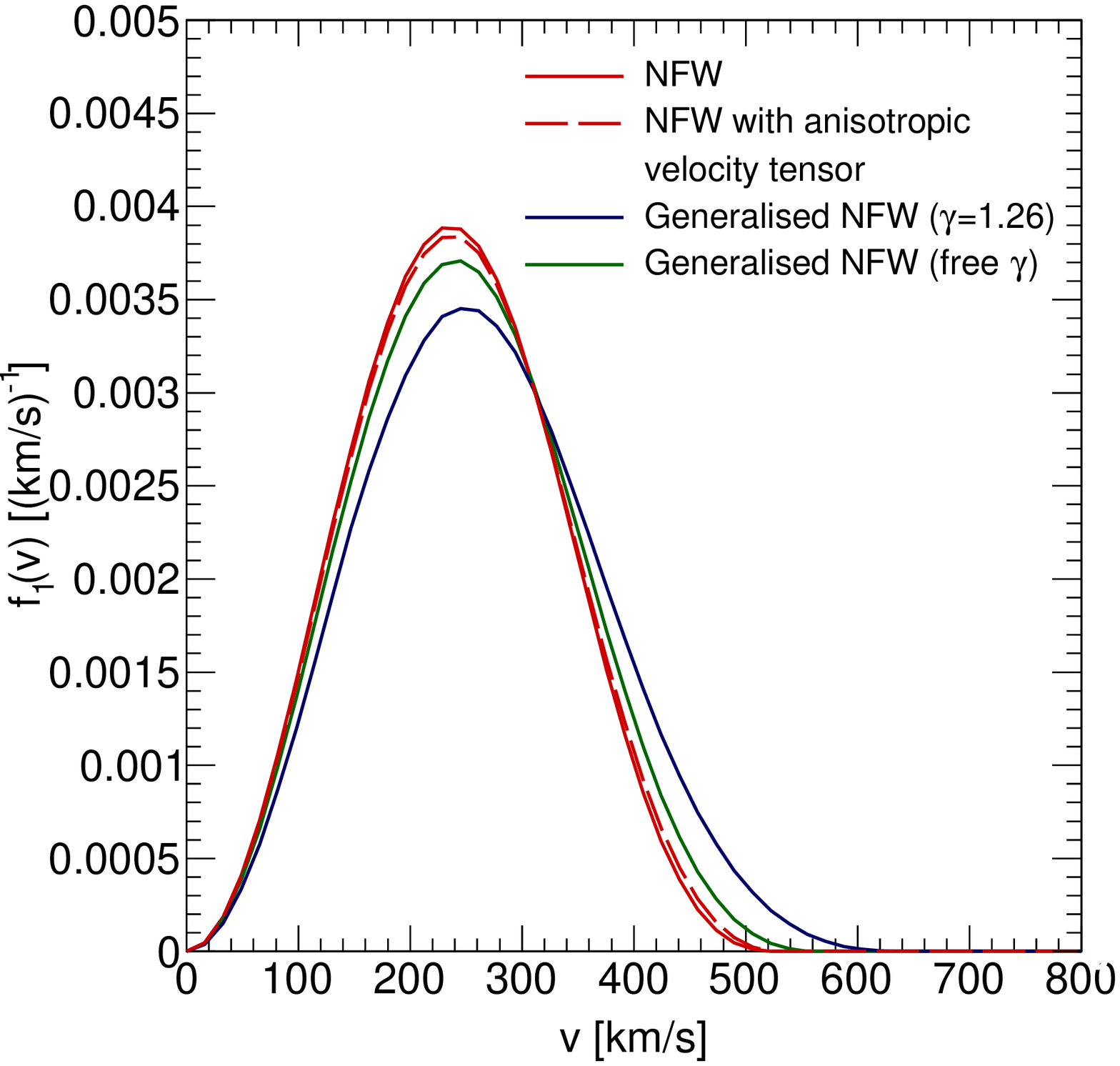}
\includegraphics[width=0.4\textwidth]{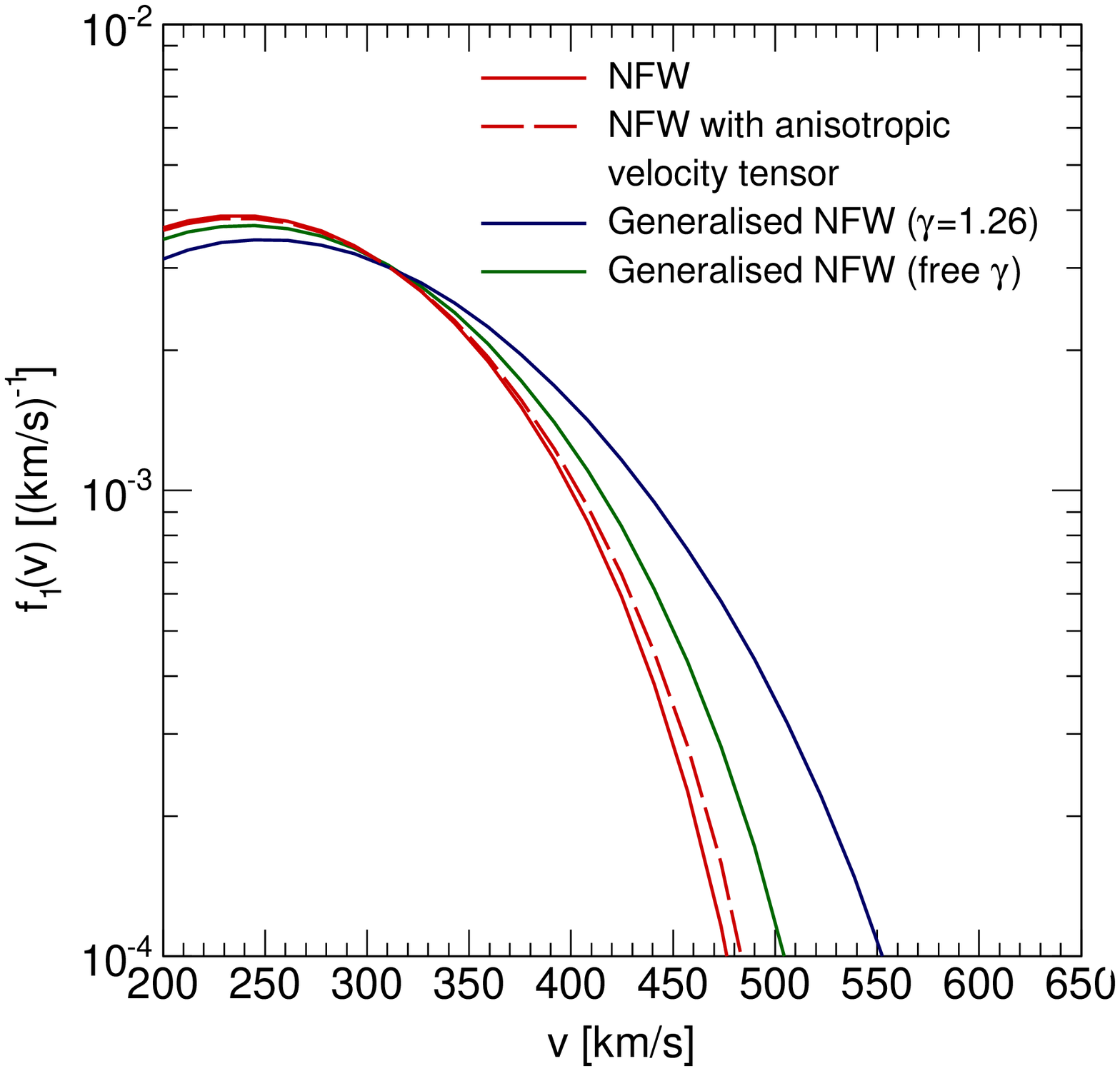}
\end{center}
\caption{\label{fig:fv_galframe} Top: local speed distribution, $f_{1}(v)$, for the ``generalised NFW (free $\gamma$)'' case. The solid green line is calculated for the best-fit point assuming isotropy, while the inner and outer contours denote the 68\% and 95\% confidence uncertainties. The dashed green line (difficult to see because it overlaps with the solid green one) is for the same best-fit point but for an anisotropic phase-space density, $F(E,L)$, as in Eq.~(\ref{eqn:F_L}). Bottom left: speed distributions for the best-fit points for the three DM halo profiles considered in our scans. The solid red line is for ``NFW'', the solid blue one for ``generalised NFW ($\gamma=1.26$)'' and the solid green one for ``generalised NFW (free $\gamma$)''. The dashed red line is for ``NFW'' but for an anisotropic phase-space density. Bottom right: same as bottom left, but for speeds larger than 200~km s$^{-1}$.  For the anisotropic velocity tensor we fix $\beta_0=0$, $\beta=0.3$ and $L_0=r_{\rm s}\Theta_0$.}
\label{fig:fv}
\end{figure*}

In the top panel of Fig.~\ref{fig:fv_galframe} we show the speed distribution and its uncertainties for the case of ``generalised NFW (free $\gamma$)'', obtained assuming an isotropic phase-space distribution, i.e. $F(E,L)=F(E)$.  The overall uncertainty on $f_1(v)$ for ``generalised NFW (free $\gamma$)'' is smaller than a factor of 2 for velocities below $440$~km s$^{-1}$. As found in Ref.~\cite{Fornasa:2013iaa}, the speed distribution has a characteristic velocity around $350$~km s$^{-1}$ where the uncertainty is smallest, which can be explained as follows. The distributions must have a peak between 200 and 300 ${\rm km \,s}^{-1}$ (in agreement with the measured value of $\Theta_0$). They are also normalized to unity, so that a large $f_1(v)$ in a given range of speed (e.g., a high peak) needs to be counterbalanced in another part of the spectrum (e.g., a depleted high-speed tail).

The bottom left panel of Fig.~\ref{fig:fv_galframe} compares the speed distributions for the best-fit points of the three DM halos considered in this work, under the hypothesis of an isotropic phase-space distribution. As expected, the position of the peak is approximately the same in each case, since it is related to $\Theta_0$. The height of the peak, however, is reduced when $\gamma$ is larger than one. We suspect that this is related to the fact that for $\gamma>1$, the DM halo is more extended than a normal NFW profile (the best-fit $r_{\rm s}$ is larger). Therefore, there are more orbits that extend to large distances from the centre, and have large speeds, and a more highly populated high-speed tail leads to a less pronounced peak. This is highlighted in the bottom right panel which focuses on the region with $v>200 \, {\rm km \, s}^{-1}$. The speed distributions for the ``NFW'' and ``generalised NFW ($\gamma=1.26$)" cases are characterised by similar uncertainty bands as the ``generalised NFW (free $\gamma$)'' case in the top panel.  For clarity, we do not display the uncertainty bands in these cases.

The dashed lines in Fig.~\ref{fig:fv_galframe} show the best-fit speed distributions for the anisotropic phase-space distribution, $F(E,L)$, for ``generalised NFW (free $\gamma$)'' (top panel), and ``NFW'' (bottom panels). We have fixed $L_0$ to be equal to $r_s \Theta_0$ and $\beta_0$ and $\beta_\infty$ to $0$ and $0.3$, respectively. These are reasonable values for MW-like halos, according to Ref.~\cite{Ludlow:2011cs}. This small deviation from isotropy has a relatively small effect on the speed distributions, and is sub-dominant to the changes that arise from varying the inner slope of the density profile. We therefore do not consider anisotropy when calculating exclusion limits in the next section.
However, Ref.~\cite{Fornasa:2013iaa} found that marginalising over $L_0$, $\beta_0$ and $\beta_\infty$ increases the size of the uncertainty band by approximately a factor of 2.

Finally, we note that the uncertainty bands in the top panel of Fig.~\ref{fig:fv_galframe}, were derived by dividing the range between 0 and 800 ${\rm km \,s}^{-1}$ into 50 bins and determining the 68\% and 95\% confidence level interval for each bin independently. As the bands are the envelope of 50 independent distributions, not all of the $f_{1}(v)$ curves that lie within them correspond to physical models for the MW. For instance, there are no models in our scans that correspond to the lower or upper edges of the uncertainty bands.

\section{LUX exclusion limits}
\label{sec:dd}
In this section we derive the upper bounds on the DM-nucleon scattering cross section, using the self-consistent speed distribution functions obtained in the previous section. We focus on the LUX experiment, making use of the 2013 published data set \cite{Akerib:2013tjd}. LUX currently provides the most stringent upper limit on the scattering cross section within the mass range favoured by a DM interpretation of the Fermi-LAT data (from approximately 8 to 200~GeV). However our procedure of using gamma-ray data to constrain the MW model can be used to calculate consistent bounds for any direct detection data set.  

The event rate in detectors based on liquid xenon is measured in terms of the prompt scintillation signal, $S1$. Following the formalism of Ref.~\cite{Aprile:2011hx}, the number of expected photoelectrons (PEs), $\nu(E_R)$, is given by
\begin{equation}
\nu(E_{\rm R}) = E_{{\rm R}} \mathcal{L}_{{\rm eff}}(E_{{\rm R}}) Q_{\gamma} \,,
\end{equation}
where $E_{\rm R}$ is the nuclear recoil energy, $Q_{\gamma}$ is the photon detection efficiency (for LUX $Q_{\gamma}=0.14$ \cite{Akerib:2013tjd}) and $\mathcal{L}_{\rm eff}(E_{{\rm R}})$ is the absolute scintillation yield, which we have digitized from Ref.~\cite{LUX_Leff:2013}. The event rate in terms of the number of PEs, $n$, is then given by
\begin{equation}
\frac{{\rm d} R}{{\rm d}n} = \int_0^\infty \frac{{\rm d} R}{{\rm d} E_{\rm R}} 
{\rm Poiss}\left( n | \nu(E_{\rm R}) \right) {\rm d} E_{\rm R}  \,,
\end{equation}
where ${\rm Poiss}\left( n | \nu(E_{\rm R}) \right)$ is a Poisson distribution with expectation value $\nu(E_{\rm R})$ and ${\rm d} R/{\rm d} E_{\rm R}$ is the usual differential event rate in terms of nuclear recoil energy (see, e.g., Ref.~\cite{Cerdeno:2010jj}). Finally, taking into account the finite average single-PE resolution of the photomultipliers, $\sigma_{\rm PMT}=0.37$~PE~\cite{Akerib:2012ys}, the resulting $S1$-spectrum is given by
\begin{equation}
\frac{{\rm d} R}{{\rm d} S1} =
\sum_{n=1}^\infty {\rm Gauss}( S1 | n ,\sqrt{n}\sigma_{\rm PMT} )  
\frac{{\rm d} R}{{\rm d} n}  \zeta(S1) \,,
\label{eq:rate}
\end{equation}
where $\zeta(S1)$ is the acceptance corresponding to the data cuts applied and is taken, in this case, from the bottom panel of Fig.~1 of Ref.~\cite{Akerib:2013tjd}, including an extra factor $1/2$ to account for the 50\% nuclear recoil acceptance. The function $ {\rm Gauss}( S1 | n ,\sqrt{n}\sigma_{\rm PMT} )$ denotes a normal distribution with mean $n$ and standard deviation $\sqrt{n}\sigma_{\rm PMT}$. We consider a range in S1 between 2 and 30 PEs and an exposure of $10\, 065.4 \, {\rm kg \, days}$.

\begin{figure*}[!t]
\begin{center}  
\includegraphics[width=0.45\textwidth]{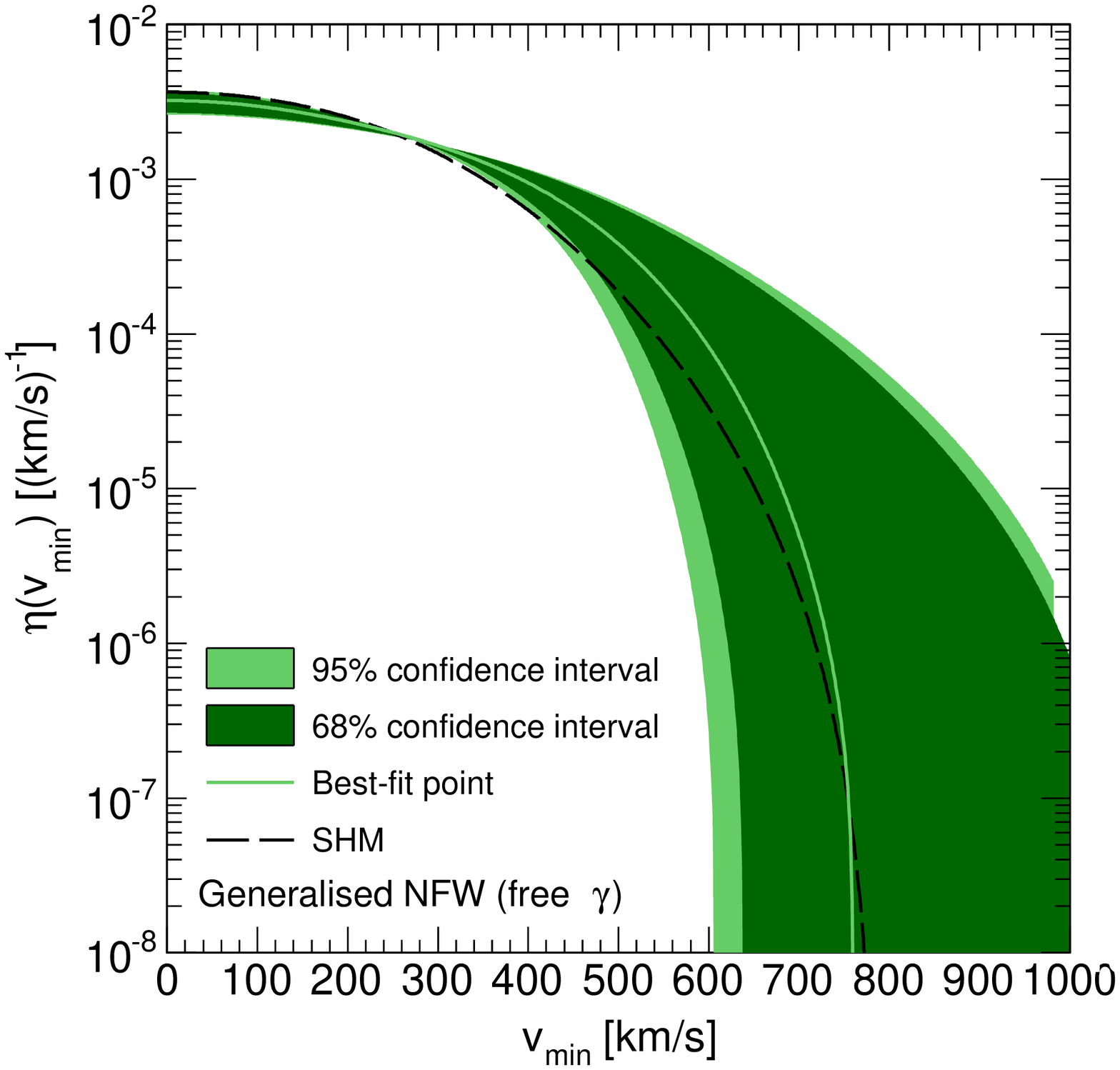}
\includegraphics[width=0.45\textwidth]{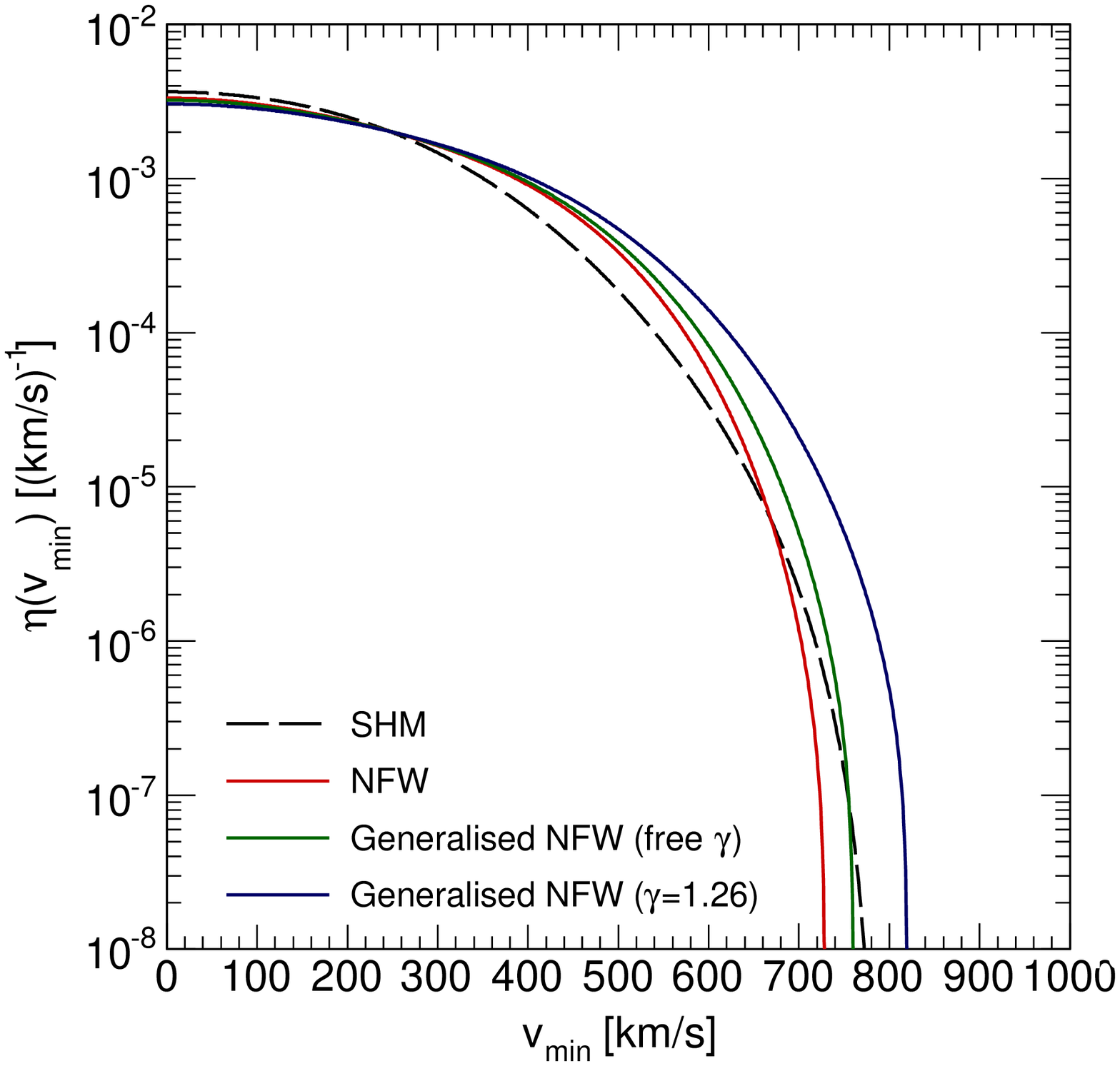}
\end{center}
\caption{Left: best-fit inverse speed function, $\eta(v_{\rm min})$, for the ``generalised NFW (free $\gamma$)'' model. The solid green line is for the best-fit point, while the inner and outer band represent the 68\% and 95\% confidence level regions. The black line is the SHM. Right: best-fit inverse speed functions for the SHM (black line), ``NFW" (red), ``generalised NFW ($\gamma=1.26$)'' (blue) and ``generalised NFW (free $\gamma$)" (green line). In all cases an isotropic phase-space density is assumed.}
\label{fig:eta}
\end{figure*}

The differential event rate, ${\rm d} R/{\rm d} E_{\rm R}$, is proportional to the mean inverse WIMP speed function,
\begin{equation}
\eta (v_{\rm min})=\int_{v_{\rm min}}\frac{\tilde{f}_1(v)}{v} \, {\rm d} v \,,
\label{eq:halo_general}
\end{equation}
where $\tilde{f}_1(v)$ is the normalised WIMP speed distribution in the detector reference frame and $v_{\rm min}$ is the minimum WIMP speed that can cause a recoil of energy $E_{\rm R}$,
\begin{equation}
v_{\rm min}  =  \left( \frac{ E_{\rm R} m_{\rm N} }{2 \mu_{{\rm N}}^2} \right)^{1/2} \,,
\end{equation}
where $m_{\rm N}$ is the atomic mass of the detector nuclei and $\mu_{{\rm N}}$ the WIMP-nuclei reduced mass. The WIMP speed distribution in the detector rest frame is calculated from the Galactic rest frame velocity distribution, $f({\bf v})$, discussed in Sec.~\ref{sec:fv}, by carrying out a Galilean transformation: ${\bf v} \rightarrow {\bf v}^{\prime} = {\bf v} + {\bf v}_{\rm e}$, where ${\bf v}_{\rm e}$ is the Earth's speed with respect to the Galactic rest frame. Ignoring the Earth's orbital speed, ${\bf v}_{\rm e}$ is the sum of the circular rotational velocity at the Solar radius, $(0 \,, \Theta_{0} \,, 0)$, and the component of the Solar peculiar velocity in the direction of Galactic rotation $V_{\odot,\phi}^{\rm RSR}$.

In the left panel of Fig.~\ref{fig:eta} we show $\eta(v_{\rm min})$ and its uncertainties for the ``generalised NFW (free $\gamma$)'' case, assuming an isotropic phase-space density. We compare it with the SHM, which has a Maxwellian speed distribution and local circular speed $\Theta_{0}=220 \, {\rm km \, s}^{-1}$, that the LUX collaboration used to derive their exclusion limits. The right panel compares $\eta(v_{\rm min})$ for the best fits of the three DM halo profiles that we have considered. The uncertainty in the three cases is similar to that for the ``generalised NFW (free $\gamma$)'' so, for clarity, we do not include it in this panel. When calculating the mean inverse speed distribution, $\Theta_0$, $v_{\rm esc}$ and $V_{\odot,\phi}^{\rm RSR}$ are needed. For consistency, in each case we use the best-fit values from the Bayesian scan performed for the corresponding halo profile~\footnote{The edges of the 68\% and 95\% uncertainty bands for $\eta(v_{\rm min})$ in the left panel of Fig.~\ref{fig:eta} are computed using the 68\% and 95\% confidence levels of the probability distributions for $\Theta_0$, $v_{\rm esc}$ and $V_{\odot,\phi}^{\rm RSR}$.}. The three self-consistent inverse speed distributions are larger than that of the SHM for $v_{\rm min} > 300 \, {\rm km \, s}^{-1}$ and the differences between them reflect the differences in the speed distributions in Fig.~\ref{fig:fv_galframe}. In particular the behaviour at very large speeds reflects their different escape speeds.
Notice that the SHM inverse mean-velocity becomes larger than that for the ``NFW" and ``Generalised NFW (free $\gamma$)" for $v_{\rm min}>700$~km~s$^{-1}$, however this only occurs when $\eta(v_{\rm min})$ is already very small.

\begin{figure*}
\begin{center}  
\includegraphics[width=0.4\textwidth]{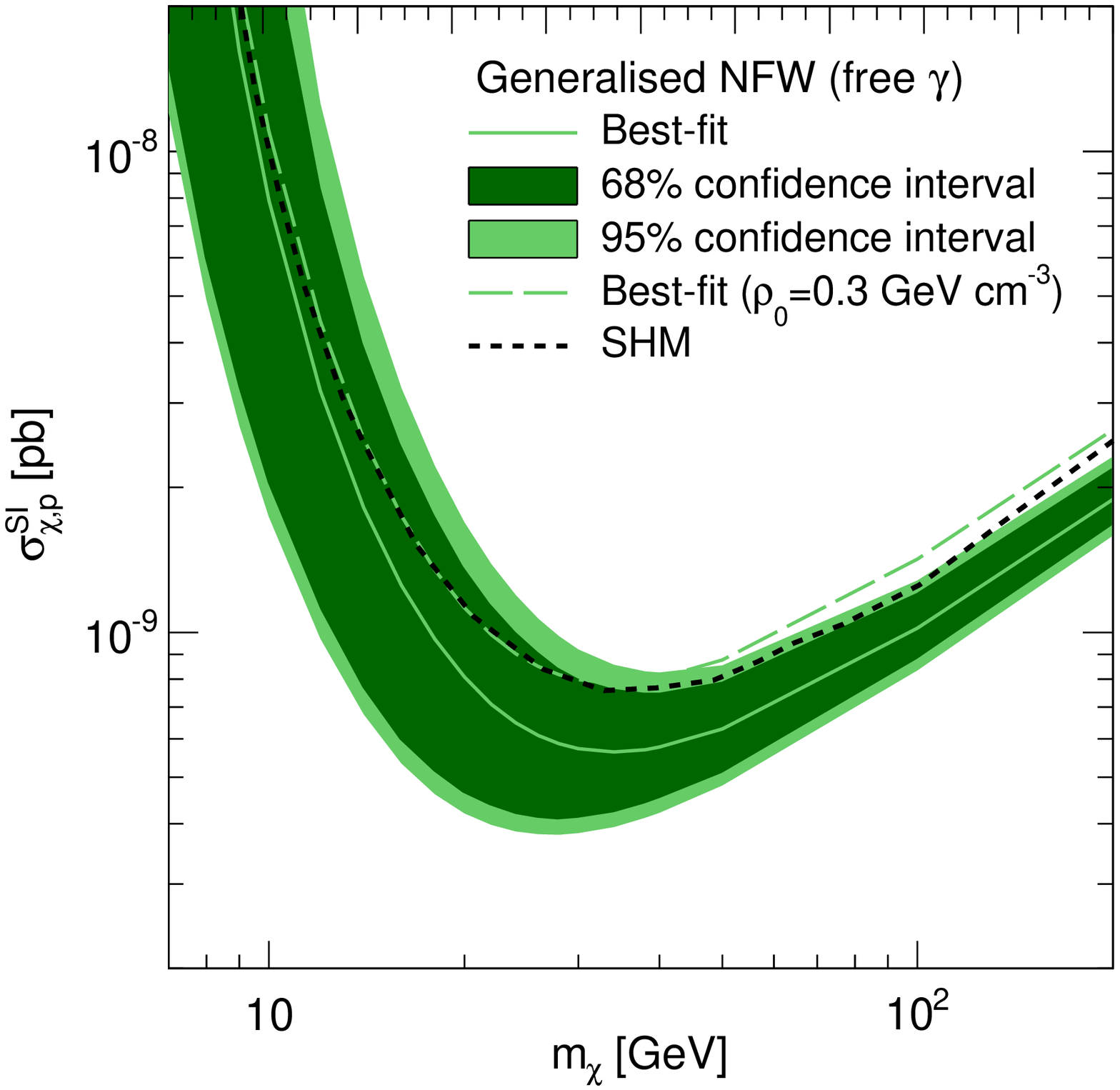}
\includegraphics[width=0.4\textwidth]{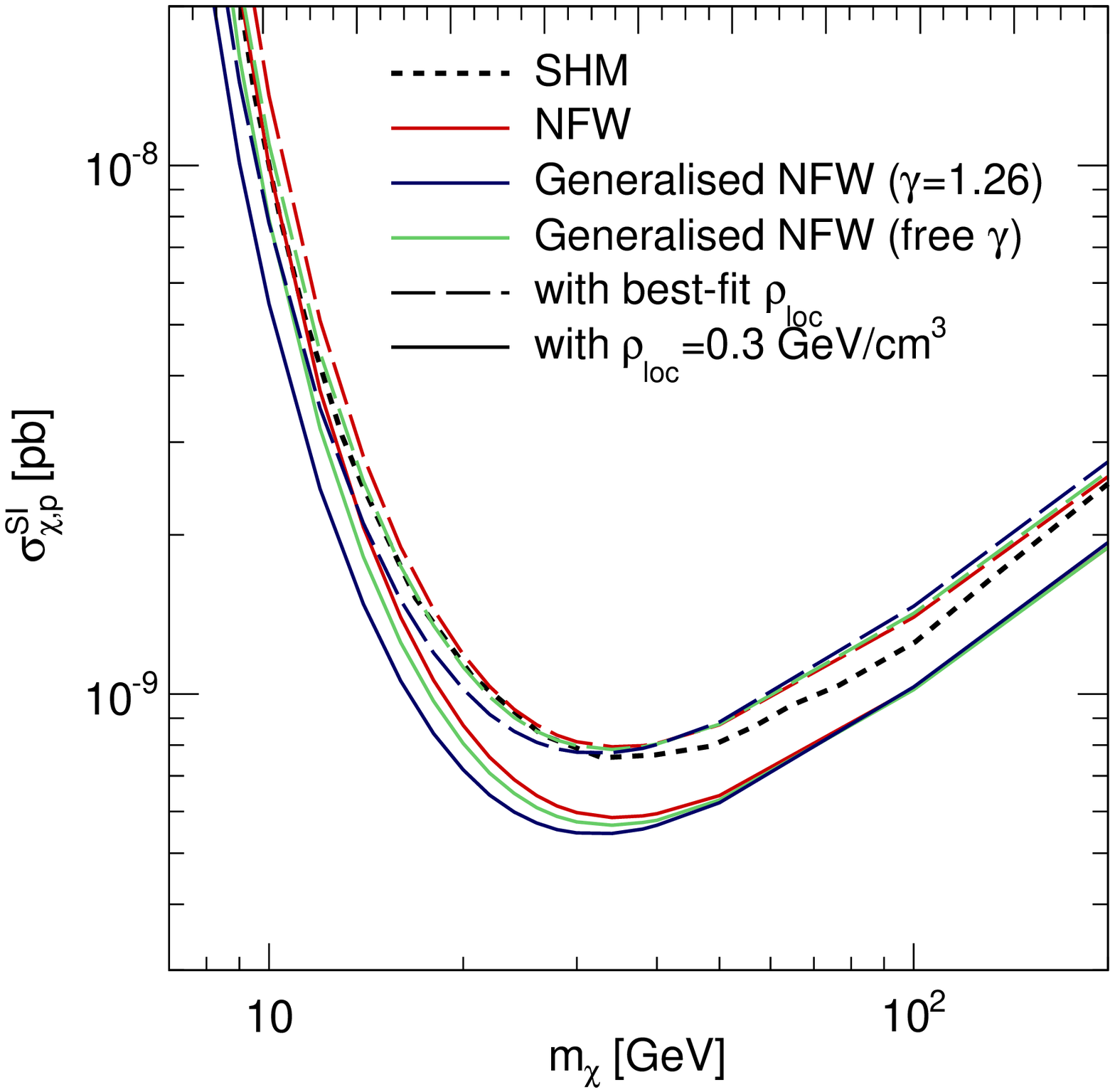}
\end{center}
\caption{Left: upper limit on the spin-independent component of the WIMP-nucleon elastic scattering cross-section, $\sigma_{\chi,{\rm p}}^{\rm SI}$ from the LUX data for the ``generalised NFW (free $\gamma$)'' case (for which the inverse speed function is shown in the left panel of Fig.~\ref{fig:eta}) and its uncertainties. The short-dashed black line corresponds to the SHM, as used by the LUX collaboration. The dashed green line uses the self-consistent speed distribution for the ``generalised NFW (free $\gamma$)'' case, but with a local density of  $\rho_0=0.3 \mbox{ GeV} \, \mbox{cm}^{-3}$. Right: exclusion limits corresponding to the best-fit solutions for each of the three density profiles considered in this analysis. The solid lines are obtained using the corresponding best-fit values of $\Theta_0$, $v_{\rm esc}$, $V_{\odot,\phi}^{\rm RSR}$ and $\rho_0$, while the dashed lines use a common local density of  $\rho_0=0.3  \mbox{ GeV} \, \mbox{cm}^{-3}$. The short-dashed black line denotes the upper limit for the SHM.}
\label{fig:ULs}
\end{figure*}

We calculate the LUX exclusion limit on the spin-independent component of the WIMP-nucleon scattering cross-section, $\sigma_{\chi,{\rm p}}^{\rm SI}$, at 90\% confidence level, using the Yellin maximum gap method~\cite{Yellin:2002xd}. The first results from LUX reported one candidate event, located at the edge of their signal region \cite{Akerib:2013tjd}. Following Ref.~\cite{Marcos:2015dza} we assume that this event is due to background leakage. The resulting exclusion limit for the SHM matches the published results of the LUX collaboration very well (see Fig. 3 of Ref.~\cite{Marcos:2015dza}). This indicates that this method is sufficient for exploring the effects of different mass models for the MW on the exclusion limit.

The expected number of recoil events is directly proportional to the product of the WIMP-nucleon scattering cross-section, $\sigma_{\chi,{\rm p}}^{\rm SI}$, and the local DM density $\rho_0$. Therefore, in order to calculate an upper limit on $\sigma_{\chi,{\rm p}}^{\rm SI}$, a value for $\rho_0$ must be specified. A change in the local DM density leads to a uniform shift in the exclusion limit for all DM masses. In the left panel of Fig.~\ref{fig:ULs}, the solid green line shows the LUX exclusion limit for the self-consistent isotropic speed distribution for the best fit of the ``generalised NFW (free $\gamma$)'' case, calculated using the best-fit density, which in this case is $\rho_0=0.42 \mbox{ GeV} \, \mbox{cm}^{-3}$ (see Ref.~\cite{Read:2014qva} for a recent review of determinations of the local DM density). Also, the short-dashed black line shows the limit obtained for the SHM, which has a local density of $\rho_0=0.3 \mbox{ GeV} \, \mbox{cm}^{-3}$. The difference between the two exclusion limits at large $m_\chi$ is around 20\% and the SHM is outside of the 68\% confidence level uncertainty of the self-consistent speed distribution for WIMP masses above $\sim 40 \, {\rm GeV}$. This discrepancy is mainly due to the different values of the local density. However, the different shapes of the exclusion limits for low WIMP masses reflect the differences in the inverse speed distributions. To emphasise this we also show (green dashed line in the left panel of Fig.~\ref{fig:ULs}) the exclusion limit using the self-consistent speed distribution but fixing the DM local density to $0.3 \mbox{ GeV} \, \mbox{cm}^{-3}$, matching that of the SHM. 

The right panel of Fig.~\ref{fig:ULs} compares the exclusion limits obtained using the best-fit solutions for the three DM density profiles we consider. As before, we also show (dashed lines) the exclusion limits using the self-consistent speed distributions and a fixed local density of $0.3 \mbox{ GeV} \, \mbox{cm}^{-3}$, rather than the best-fit values, in order to isolate the effect of the speed distribution from that of the density. The differences in the speed distributions mainly affect the exclusion limits in the low-mass regime, 
which differ from that of the SHM by tens of per-cent
for $m_\chi\lesssim 40$~GeV. 
Depending on the halo profile, the limits can be either tighter (for the generalised NFW with $\gamma=1.26$) or weaker (NFW and generalised NFW with free $\gamma$).

As discussed above, and illustrated in Fig.~\ref{fig:fv_galframe}, for fixed values of $\beta_0$, $\beta_\infty$ and $L_0$ that are reasonable for the MW halo, anisotropy in the velocity tensor affects the speed distribution less than changing the inner slope of the density profile. We therefore do not show the exclusion limits for the best-fit anisotropic cases in Fig.~\ref{fig:ULs}. Note however that marginalising over the anisotropy parameters significantly increases the uncertainty in the speed distribution, and hence also the uncertainty in the exclusion limits.

\section{Conclusions}
\label{sec:discussion}
When comparing results from direct and indirect WIMP searches it is crucial to ensure that the assumptions made about the properties of the DM halo are consistent. For example, the constraints on the WIMP mass and scattering cross-section off quarks from direct detection experiments are sensitive to the local DM density and velocity distribution, both of which depend on the DM density profile. On the other hand the morphology of the gamma-ray flux from WIMP annihilation in the MW halo would constrain the DM density profile. To illustrate how to consistently include such information in the calculation of direct detection bounds, we investigate the case of the Fermi-LAT gamma-ray excess from the Galactic Centre being due to WIMP annihilation.

We have derived limits on the spin-independent WIMP-nucleon elastic scattering cross section from the LUX data, for best-fit MW mass models that are compatible with a DM interpretation of the Galactic Centre excess. We did this in two different ways. Firstly we enforced compatibility by simply fixing the logarithmic slope of the DM halo density profile to the value which gives the best fit to the morphology of the Galactic Centre excess, $\gamma=1.26$. Secondly we allowed the inner slope to vary and included the morphology of the Galactic Centre excess in the data sets used to constrain the mass model. In each case we used multiple observational data sets, such as stellar kinematics and microlensing, to constrain the parameters of the mass model. We found that DM halos with an inner slope $\gamma$ larger than 1 tend to be more extended, having a larger scale radius $r_{\rm s}$ and a lower concentration. On the contrary, the baryonic component is well constrained by observation in the inner Galaxy and from local measurement. Thus, its properties do not change significantly when the GeV excess is included in the observational constraints.

We then derived the DM speed distribution from the inferred gravitational potential of the MW in a self-consistent way, in the case an isotropic phase-space density, $F(E,L)=F(E)$. These self-consistent speed distributions have more particles in their high-speed tails than the SHM, which is probably related to the fact that the best-fit DM halo is more extended than the SHM. Also, we have found that this has a significant effect on the cross-section exclusion limit from the LUX data. As previously found in e.g. Ref.~\cite{Catena:2009mf,McMillan:2011wd}, the best-fit MW mass models have values for the local DM density that are larger than that of the SHM, which tightens the exclusion limit for all WIMP masses. The self-consistent speed distributions affect the exclusion limits for low masses, e.g. below 40 GeV. For these light WIMPs, the total number of events expected in LUX is found by integrating the inverse speed distribution $\eta(v_{\rm min})$ above a $v_{\rm min}$ of, approximately, $200-300$~km s$^{-1}$ (see Eq.~\ref{eq:halo_general}). As discussed in Sec.~\ref{sec:fv} (and shown in Figs.~\ref{fig:fv} and \ref{fig:eta}), the high-speed tail of the speed distribution is more populated for the generalised NFW halo profiles. The ``generalised NFW ($\gamma$=1.26)'' case yields the largest number of expected events, followed by ``generalised NFW (free $\gamma$)''. Therefore the exclusion limits for light WIMPs are tightest for the ``generalised NFW ($\gamma$=1.26)'' case, and both of the generalised NFW models produce tighter exclusion limits than the standard NFW profile. 
For light WIMPs the exclusion limits for the self-consistent velocity distributions differ from that for the SHM by tens of per-cent and, depending on the halo profile, can be either tighter or weaker.

We also considered the possibility of a DM halo with an anisotropic velocity tensor, parameterising the $L$-dependent component of the phase-space density according to Ref.~\cite{Wojtak:2008mg}. For reasonable, fixed values of the anisotropy parameters, the speed distributions are very similar to the isotropic case (for a given DM density profile). Therefore we expect the LUX exclusion limits to be very similar to the isotropic case. Marginalising over the unknown anisotropy parameters would, however, significantly increase the uncertainty on both the speed distribution and the exclusion limits.

Our work reinforces the need for a consistent interpretation of data from DM experiments. Combining data from different strategies allows not only an improved reconstruction of the properties of DM, but also better control over the uncertainties involved. The amount of experimental data on DM is rapidly increasing, alongside the precision of theoretical models. It is, therefore, vital to identify and apply good practice in the way different data sets are combined.

\vspace*{1cm}
\acknowledgments

D.G.C. is funded by the STFC. M.F. gratefully acknowledges support from the Netherlands Organization for Scientific Research (NWO) through a Vidi grant (P.I.: Dr. Shin'ichiro Ando). A.M.G. acknowledges  support  from  STFC  grant ST/L000393/1. M.P. is supported under the ERC Advanced Grant SPLE under contract ERC-2012-ADG-20120216-320421. We acknowledge support of the Consolider-Ingenio 2010 program under grant MULTIDARK CSD2009-00064, the Spanish MICINN under Grant No. FPA2015-65929-P and the Spanish MINECO Centro de Excelencia Severo Ochoa Program under Grant No. SEV-2012-0249.

\end{document}